\documentclass[runningheads]{llncs}
\usepackage[T1]{fontenc}
\usepackage{graphicx}
\usepackage{marvosym}
\usepackage{amsmath}
\usepackage{amssymb}
\usepackage{pdfsync}

\begin{document}
\title{Constrained Heterogeneous Two-facility Location Games with Max-variant Cost}
%
%
\author{Qi Zhao \and
Wenjing Liu\orcidID{0000-0003-4826-2088}\thanks{The corresponding author.} \and
Qizhi Fang \and
Qingqin Nong\orcidID{0000-0002-0895-7793}}
\authorrunning{Q. Zhao et al.}

\institute{Ocean University of China, Qingdao 266100, China\\
\email{zq1012@stu.ouc.edu.cn}, \email{liuwj@ouc.edu.cn}, \email{qfang@ouc.edu.cn}, \email{qqnong@ouc.edu.cn}}

\maketitle              
\begin{abstract}
In this paper, we propose a constrained heterogeneous facility location model where a set of alternative locations are feasible for building facilities and the number of facilities built at each location is limited. Supposing that a set of agents on the real line can strategically report their locations and each agent's cost is her distance to the further facility that she is interested in, we study deterministic mechanism design without money for constrained heterogeneous two-facility location games.

Depending on whether agents have optional preference, the problem is considered in two settings: the compulsory setting and the optional setting. In the compulsory setting where each agent is served by the two heterogeneous facilities, we provide a 3-approximate deterministic group strategyproof mechanism for the sum/maximum cost objective respectively, which is also the best deterministic strategyproof mechanism under the corresponding social objective. In the optional setting where each agent can be interested in one of the two facilities or both, we propose a deterministic group strategyproof mechanism with approximation ratio of at most $2n+1$ for the sum cost objective and a deterministic group strategyproof mechanism with approximation ratio of at most 9 for the maximum cost objective.

\keywords{Mechanism design \and Facility location \and Strategyproof \and Constrained}
\end{abstract}

\section{Introduction}

In the origin mechanism design problem for heterogeneous facility location games, there are a set of strategic agents who are required to report their private information and a social planner intends to locate several heterogeneous facilities by a mechanism based on the reported information, with the purpose of optimizing some social objective. In this paper, we study the problem of locating two heterogeneous facilities under a constrained setting, which means a set of alternative locations are feasible for building facilities and the number of facilities built at each location is limited.

Compared with the origin setting where facilities can be built anywhere in a specific metric space and there is no limit on the number of facilities at each location, our constrained setting models well many practical applications. For example, in the realistic urban planning, facilities can only be built at designated sites and the number of facilities at each site is limited. To accommodate these constraints, we propose a multiset of feasible locations and at most one facility is permitted to build at each location. Further, we focus on the Max-variant where the cost of each agent depends on her distance to the farthest one if she is served by two or more heterogeneous facilities. The Max-variant can be found applications in natural scenarios~\cite{Yuan2016}. For example, a local authority plans to locate different raw material warehouses for several processing plants. Assuming each plant has multiple transport trucks having the same speed, the time that the plant has to wait depends on its distance to the farthest one if it requires raw materials from different sites.

We discuss the mechanism design problem for constrained heterogeneous two-facility location games with Max-variant cost in two settings: the first is the compulsory setting, where each agent is served by the two heterogeneous facilities; the second is the optional setting, where each agent is served by either one of the two facilities or both. Considering that agents may manipulate the facility locations by misreporting their private information, we concentrate on mechanisms that can perform well under some social objective (e.g., minimizing the sum/maximum cost) while guaranteeing truthful report from agents (i.e., strategyproof or group strategyproof).

\subsection{Our Contribution}

This paper studies deterministic mechanism design without money for constrained heterogeneous two-facility location games with Max-variant cost under the objective of minimizing the sum/maximum cost.

Our key innovations and results are summarized as follows.

In Section \ref{model}, we formulate the constrained heterogeneous facility location game with Max-variant cost. We propose a finite multiset of alternative locations which are feasible for building facilities and require that at most one facility can be built at each location. Thus, by adjusting the number of same elements in the multiset, the model can accommodate different scenarios where the number of facilities at the same location is limited.

In Section \ref{compulsory}, we focus on deterministic mechanism design in the compulsory setting. We propose a set of adjacent alternative location pairs, which all agents have single peaked preferences over and the optimal solution under the sum/maximum cost objective can always be found in. We prove that any deterministic strategyproof mechanism has an approximation ratio of at least 3 under the sum/maximum cost objective. In addition, we present 3-approximate deterministic group strategyproof mechanisms for both social objectives, which implies that the best deterministic strategyproof mechanisms have been obtained.

In Section \ref{optional}, we discuss the optional setting. For the sum cost objective, we propose a deterministic group strategyproof mechanism with approximation ratio of at most $2n+1$. For the maximum cost objective, we design a deterministic group strategyproof mechanism with approximation ratio of at most 9.


\subsection{Related Work}\label{related}

Mechanism design without money for facility location games has been extensively studied in recent years.
Early studies focused on the characterization of strategyproof mechanisms. Moulin~\cite{Moulin1980} identified all the possible strategyproof mechanisms for one-facility location on the line with single peaked preferences, whose results were extended by Schummer \& Vohra~\cite{Schummer2002} and Dokow et al.~\cite{Dokow2012} to tree and cycle networks.

Approximate mechanism design without money was initiated by Procaccia \& Tennenholtz~\cite{Procaccia2009}, who studied deterministic and randomized strategyproof mechanisms with constant approximation ratio for facility location games under the sum cost and the maximum cost in three settings: one-facility, two-facility and multiple facilities per agent. Following this research agenda, numerous studies have emerged, including improvements on the lower/upper bound of approximation~\cite{Lu2010,Fotakis2014} and further variants.

Cheng et al.~\cite{Cheng2013} introduced approximate mechanism design for obnoxious facility location games where the facility is not desirable to each agent.
Zou \& Li~\cite{Zou2015} studied the dual preference setting where the facility can be desirable or undesirable for different agents.
Zhang \& Li~\cite{Zhang2014} introduced weights to agents and Filos-Ratsikas et al.~\cite{Filos2017} studied one-facility location problem with double-peaked preferences.
Serafino \& Ventre~\cite{SV2016} introduced heterogeneous two-facility location games where each agent cares about either one facility or both and her cost depends on the sum of distances to her interested facilities (referred to as the Sum-variant).
Later, Yuan et al.~\cite{Yuan2016} considered the Min-variant and Max-variant instead and Anastasiadis \& Deligkas~\cite{Anastasiadis2018} studied heterogeneous $k$-facility setting with Min-variant.
Besides, various individual and social objectives were also studied. Mei et al.~\cite{Mei2019} introduced a happiness factor to measure each agent's individual utility. Feigenbaum \& Sethuraman~\cite{Feigenbaum2017} considered the $L_p$-form of the vector of agent-costs instead of the classic sum cost. Cai et al.~\cite{Cai2016} and Chen et al.~\cite{ChenF2021} studied facility location problems under the objective of minimizing the maximum envy. Ding et al.~\cite{Ding2020} and Liu et al.~\cite{Liu2021} considered the envy ratio objective. Zhou et al.~\cite{Zhou2021} studied group-fair facility location problems.

Further, motivated by real-world applications, researchers have begun to study the mechanism design problem with constraints on the facilities. Aziz et al.~\cite{Aziz2020a,Aziz2020b} studied facility location problems with capacity constraints. Chen et al.~\cite{ChenH2021} studied the two-opposite-facility location problem with maximum distance constraint by imposing a penalty. Xu et al.~\cite{Xu2021} studied minimum distance requirement for the heterogeneous two-facility location problem.
In addition, considering that in reality the feasible locations that facilities could be built at are usually limited, mechanism design for facility location games with limited locations were also studied.
Sui \& Boutilier~\cite{Sui2015} studied approximately strategyproof mechanisms for facility location games with constraints on the feasible placement of facilities. Feldman et al.~\cite{Feldman2016} studied the one-facility location setting under the sum cost objective in the context of voting embedded in some underlying metric space. Tang et al.~\cite{Tang2020} further considered the maximum cost objective and the two-facility setting. Li et al.~\cite{Li2020} studied the heterogeneous two-facility setting with optional preference, which is also the most related to our work among all studies on the constrained heterogeneous facility location problem. However, there are at least three differences between us: (1) our model requires a limit on the number of facilities at each feasible location and \cite{Li2020} does not;
(2) each agent's location is private and her preference on facilities is public in our model while it is the opposite in \cite{Li2020};
(3) we consider the Max-variant cost while \cite{Li2020} considers the Min-variant where the cost of each agent depends on her distance to the closest facility within her acceptable set.

\section{Model}\label{model}

Let $N=\{1,2, \ldots, n\}$ be a set of agents located on the real line $\mathcal{R}$ and $\mathcal{F}=$ $\left\{F_{1}, F_{2}\right\}$ be the set of two heterogeneous facilities to be built.
Each agent $i\in N$ has a location $x_{i} \in \mathcal{R}$ and a facility preference $p_{i} \subseteq \mathcal{F}$, where $x_{i}$ is $i$'s private information and $p_{i}$ is public.
Denote $\mathbf{x}=\left(x_{1}, x_{2}, \ldots, x_{n}\right)$ and $\mathbf{p}=\left(p_{1}, p_{2}, \ldots, p_{n}\right)$ as the $n$ agents' location profile and facility preference profile, respectively.
For $i\in N$, let $\mathbf{x}_{-i}=(x_1,\ldots,x_{i-1},x_{i+1},\ldots,x_n)$ be the location profile without agent $i$, then $\mathbf{x}=(x_i,\mathbf{x}_{-i})$.
For $S\subseteq N$, denote $\mathbf{x}_S=\left(x_i\right)_{i\in S}$, $\mathbf{p}_S=\left(p_i\right)_{i\in S}$, and $\mathbf{x}_{-S}=\left(x_i\right)_{i\notin S}$, then $\mathbf{x}=\left(\mathbf{x}_S,\mathbf{x}_{-S}\right)$.

Let $A=\{a_1, a_2, \ldots, a_m\}\in \mathcal{R}^m$ be a multiset of alternative locations which are feasible for building facilities and at most one facility can be built at each location. Assume without loss of generality that $a_1\le a_2\le \ldots\le a_m$. Denote an instance of the $n$ agents by $I(\mathbf{x},\mathbf{p},A)$ or simply by $I$ without confusion.

\vspace{1.2ex}
\noindent\textbf{Individual and Social Objectives.}
When locating $F_1, F_2$ at $y_1\in A, y_2\in A\backslash\{y_1\}$ respectively, denote the facility location profile by $\mathbf{y}=(y_1, y_2)$.  Under Max-variant, the cost of agent $i$ is denoted by $c_i(\mathbf{y},(x_i,p_i))=\max_{F_j\in p_i}|y_j-x_i|$.
While each agent seeks to minimize her individual cost, the social planner aims to minimize the sum cost or maximum cost of the $n$ agents.
For a location and facility preference profile $(\mathbf{x},\mathbf{p})\in \mathcal{R}^n\times \left(2^{\mathcal{F}}\right)^n$, the sum cost and the maximum cost under $\mathbf{y}$ are denoted by $sc(\mathbf{y},(\mathbf{x},\mathbf{p}))=\sum_{i\in N}c_i(\mathbf{y},(x_i,p_i))$ and $mc(\mathbf{y},(\mathbf{x},\mathbf{p}))=\max_{i\in N}c_i(\mathbf{y},(x_i,p_i))$, respectively.
Let $OPT_{sc}(\mathbf{x},\mathbf{p})$ and $OPT_{mc}(\mathbf{x},\mathbf{p})$ be the optimal solution under the sum cost and the maximum cost, respectively.

Considering the limit on facility locations, the mechanism in our constrained setting is defined as follows.

\begin{definition}
A \textbf{deterministic mechanism} $f$ is a function that maps the $n$ agents' location profile $\mathbf{x}$ and facility preference profile $\mathbf{p}$ to a location profile of the two facilities, i.e., $f(\mathbf{x},\mathbf{p})=\mathbf{y}=(y_1, y_2), \forall (\mathbf{x},\mathbf{p})\in \mathcal{R}^n\times \left(2^{\mathcal{F}}\right)^n$, where $\mathbf{y}=(y_1, y_2)$ should satisfy $y_1\in A$ and $y_2\in A\backslash\{y_1\}$.
\end{definition}

Given a mechanism $f$ and a reported location profile $\mathbf{x}'\in \mathcal{R}^n$, the cost of agent $i\in N$ under $f$ is $c_i(f(\mathbf{x}',\mathbf{p}),(x_i,p_i))$.
The sum cost and maximum cost of $f$ are $sc(f(\mathbf{x}',\mathbf{p}),(\mathbf{x},\mathbf{p}))=\sum_{i\in N}c_i(f(\mathbf{x}',\mathbf{p}),(x_i,p_i))$ and $mc(f(\mathbf{x}',\mathbf{p}),(\mathbf{x},\mathbf{p}))=\max_{i\in N}c_i(f(\mathbf{x}',\mathbf{p}),(x_i,p_i))$, respectively.
Since agents may misreport their locations to benefit themselves, strategyproofness of mechanisms becomes necessary.

\begin{definition}
A mechanism $f$ is \textbf{strategyproof} if each agent can never benefit from misreporting her location, regardless of the others' strategies, i.e., for every location and facility preference profile $(\mathbf{x},\mathbf{p})\in \mathcal{R}^n\times \left(2^{\mathcal{F}}\right)^n$, every agent $ i \in N$, and every $x_i'\in\mathcal{R}$,
$c_i(f(\mathbf{x},\mathbf{p}),(x_i,p_i))\le c_i(f((x_i',\mathbf{x}_{-i}),\mathbf{p}),(x_i,p_i))$.
\end{definition}

\begin{definition}
A mechanism $f$ is \textbf{group strategyproof} if for any group of agents misreporting their locations, at least one of them cannot benefit regardless of the others' strategies, i.e., for every location and facility preference profile $(\mathbf{x},\mathbf{p})\in \mathcal{R}^n\times \left(2^{\mathcal{F}}\right)^n$, every group of agents $S\subseteq N$ and every $\mathbf{x}_S'\in\mathcal{R}^{|S|}$, there exists $i\in S$ such that
$c_i(f(\mathbf{x},\mathbf{p}),(x_i,p_i))\le c_i(f((\mathbf{x}_S',\mathbf{x}_{-S}),\mathbf{p}),(x_i,p_i))$.
\end{definition}

We aim at deterministic strategyproof or group strategyproof mechanisms that can perform well under the sum/maximum cost objective. The worst-case approximation ratio is used to evaluate a mechanism's performance.
Without confusion, denote $sc(f(\mathbf{x}',\mathbf{p}),(\mathbf{x},\mathbf{p}))$, $sc\left(OPT_{sc}(\mathbf{x},\mathbf{p}),(\mathbf{x},\mathbf{p})\right)$, $mc(f(\mathbf{x}',\mathbf{p}),(\mathbf{x},\mathbf{p}))$ and $mc\left(OPT_{mc}(\mathbf{x},\mathbf{p}),(\mathbf{x},\mathbf{p})\right)$ by
$sc(f,(\mathbf{x},\mathbf{p}))$, $sc\left(OPT,(\mathbf{x},\mathbf{p})\right)$,
$mc(f,(\mathbf{x},\mathbf{p}))$ and $mc\left(OPT,(\mathbf{x},\mathbf{p})\right)$ respectively for simplicity.
The approximation ratio under the sum cost objective is defined as follows and it is similar under the maximum cost objective.

\begin{definition}
A mechanism $f$ is said to have an \textbf{approximation ratio} of $\rho(\rho\ge 1)$ under the sum cost objective, if
\begin{equation}
\rho=\sup_{I(\mathbf{x},\mathbf{p},A)}\frac{sc(f,(\mathbf{x},\mathbf{p}))}{sc(OPT,(\mathbf{x},\mathbf{p}))}.
\end{equation}
\end{definition}

In this paper, we are interested in deterministic strategyproof or group strategyproof mechanisms with small approximation ratio under the sum/maximum cost objective.

\vspace{1.2ex}
\noindent\textbf{Notations.}
For a location profile $\mathbf{x} \in \mathcal{R}^n$, denote the median location in $\mathbf{x}$ by $\mathrm{med}$$(\mathbf{x})$, the leftmost location in $\mathbf{x}$ by $\mathrm{lt}(\mathbf{x})=\min _{i \in N} \{x_{i}\}$, the rightmost location by $\mathrm{rt}$$(\mathbf{x})=\max _{i \in N} \{x_{i}\}$, and the center location by $\mathrm{cen}(\mathbf{x})=\frac{\mathrm{lt}(\mathbf{x})+\mathrm{rt}(\mathbf{x})}{2}$.
For a facility preference profile $\mathbf{p}\in\left(2^{\mathcal{F}}\right)^n$, denote $N_{k}=\left\{i\in N\mid p_{i}=\left\{F_{k}\right\}\right\}$ for $k\in\{1,2\}$, and $N_{1,2}=\left\{i \in N\mid p_{i}=\left\{F_{1}, F_{2}\right\} \right\}$.

\section{Compulsory Setting}\label{compulsory}

In this section, we study the compulsory setting where each agent is served by the two heterogeneous facilities, i.e., $p_i=\{F_1,F_2\}, \forall i\in N$. For simplicity, we omit $p_i$ or $\mathbf{p}$ in this section. For example, replace $(\mathbf{x},\mathbf{p})$ by $\mathbf{x}$ and the cost of agent $i\in N$ under the facility location profile $\mathbf{y}=(y_1,y_2)$ is denoted by $c_i(\mathbf{y},x_i)=\max_{j\in\{1,2\}}|y_j-x_i|$.

For the multiset of alternative locations $A=\{a_1,\ldots,a_m\}$ with $a_1\le\ldots\le a_m$, denote  $AP=\{(a_1,a_2),(a_2,a_3),\ldots,(a_{m-1},a_m)\}$. Then the real line can be partitioned into $m-1$ zones where the $k$th zone (denoted by $Z_k, k=1,\ldots,m-1$) represents the set of points whose favorite location pair in $AP$ is $(a_k,a_{k+1})$. We refer to $Z_k$ as the zone of location pair $(a_k,a_{k+1})$. Obviously, it holds that
\begin{equation}
Z_k=\left\{
\begin{array}{ll}
\left(-\infty,\frac{ a_k+a_{k+2}}{ 2}\right],&k=1\\
\left(\frac{a_{k-1}+a_{k+1}}{2},\frac{a_{k}+a_{k+2}}{2}\right],&2\le k\le m-2\\
\left(\frac{a_{k-1}+a_{k+1}}{2},+\infty\right),&k=m-1
\end{array}\right.
\end{equation}

The preferences of all agents over $AP$ are (not strictly) \textit{single peaked}: for each agent $i\in N$ with location $x_i\in Z_l$, her \textit{peak} (or favorite) in $AP$ is $(a_l,a_{l+1})$ and her cost under $(a_k,a_{k+1})$ monotonically increases as $|k-l|$ increases. Based on the single peaked preference, locating at the peak of $\mathbf{x}$'s any $i$th statistic order (denoted by $\mathbf{x}_{(i)}$) is group strategyproof.

\begin{lemma}\label{gsp}
Given a location profile $\mathbf{x}$, locating at the peak of $\mathbf{x}_{(i)}$ in $AP$ for any $i\in\{1,2,\ldots, n\}$ is group strategyproof.
\end{lemma}

Lemma \ref{gsp} provides a class of group strategyproof mechanisms for the compulsory setting where all agents are served by two facilities. Next we will select proper mechanisms from this class for the sum/maximum cost objective respectively.

\subsection{Sum Cost}

For the sum cost objective, we first show that there exists an optimal solution where the two facilities are located at adjacent alternatives.

\begin{lemma}\label{opt-sc}
Given a location profile $\mathbf{x}\in\mathcal{R}^n$, there exists an optimal solution in $AP$ under the sum cost objective.
\end{lemma}

Intuitively, each agent always prefers the two facilities located as close as possible, since her cost depends on her distance to the farther one.
By Lemma \ref{opt-sc}, an optimal solution (or mechanism) can always be found in $m-1$ steps. However, it may be not strategyproof.
Consider an instance $I(\mathbf{x},A)$ with $\mathbf{x}=(0,2), A=\{-1-2\varepsilon, -1, 1+3\varepsilon\}$ where $\varepsilon>0$ is sufficiently small. It holds that $OPT_{sc}(\mathbf{x})=(-1, 1+3\varepsilon), c_1(OPT_{sc}(\mathbf{x}),x_1)=1+3\varepsilon$.
Replacing $x_1=0$ by $x_1'=-1$, we have $OPT_{sc}(\mathbf{x}')=(-1-2\varepsilon,-1), c_1(OPT_{sc}(\mathbf{x}'),x_1)=1+2\varepsilon$. Thus, agent 1 with $x_1=0$ can strictly decrease her cost by reporting $x_{1}'=-1$.

\begin{theorem}\label{sc-lower}
Under the sum cost objective, any deterministic strategyproof mechanism has an approximation ratio of at least 3.
\end{theorem}

\noindent\textit{Mechanism 1.}
Given a location profile $\mathbf{x}\in\mathcal{R}^n$, output the peak of $\mathrm{med}(\mathbf{x})$ in $AP$, i.e., the location pair $(y_1, y_2)\in\underset{(s_1,s_2)\in AP}{\arg \min } \max_{j\in\{1,2\}} |s_j-\mathrm{med}(\mathbf{x})|$, breaking ties in any deterministic way.

\begin{theorem}\label{sc-upper}
Mechanism 1 is group strategyproof and has an approximation ratio of 3 under the sum cost objective.
\end{theorem}

\subsection{Maximum Cost}

Compared with the sum cost objective, there is a more precise statement on the optimal solution under the maximum cost objective.

\begin{lemma}\label{opt-mc}
Given a location profile $\mathbf{x}\in \mathcal{R}^n$, the peak of $\mathrm{cen}(\mathbf{x})$ in $AP$ is exactly an optimal solution under the maximum cost objective.
\end{lemma}

However, the optimal mechanism is not strategyproof. Consider an instance $I(\mathbf{x},A)$ with $\mathbf{x}=(-\varepsilon, \varepsilon)$ and $A=\{-1,1,1+\varepsilon\}$. It holds that $OPT_{mc}=(-1,1)$ and $c_2(OPT(\mathbf{x}),x_2)=1+\varepsilon$ for sufficiently small $\varepsilon>0$. Replacing $x_2=\varepsilon$ by $x_2'=2$, we have $OPT_{sc}(\mathbf{x}')=(1, 1+\varepsilon), c_2(OPT_{sc}(\mathbf{x}'),x_2)=1$. Thus, agent 2 with $x_2=\varepsilon$ can strictly decrease her cost by misreporting $x_{2}'=2$.

\begin{theorem}\label{mc-lower}
Under the maximum cost objective, any deterministic strategyproof mechanism has an approximation ratio of at least 3.
\end{theorem}

\noindent\textit{Mechanism 2.}
Given a location profile $\mathbf{x}\in\mathcal{R}^n$, output the peak of $\mathrm{lt}(\mathbf{x})$ in $AP$, i.e., the location pair $(y_1, y_2)\in\underset{(s_1,s_2)\in AP}{\arg \min}\max_{j\in\{1,2\}} |s_j-\mathrm{lt}(\mathbf{x})|$, breaking ties in any deterministic way.

\begin{theorem}\label{mc-upper}
Mechanism 2 is group strategy-proof and has an approximation ratio of 3 under the maximum cost objective.
\end{theorem}

\section{Optional Setting}\label{optional}

In this section, we discuss the optional setting where each agent can be interested in either one of the two heterogeneous facilities or both. The cost of agent $i\in N$ is $c_i(\mathbf{y},(x_i,p_i))=\max_{F_k\in p_i}|y_k-x_i|$.

Note that even in the optional setting, each agent $i\in N$ has some kind of single peaked preference: if $p_i=\{F_1\}$ or $\{F_2\}$, she has single peaked preference over $A$; if $p_i=\{F_1,F_2\}$, she has single peaked preference over $AP$. Our mechanisms will be proposed based on the single peaked preference.

In the following subsections, two mechanisms for one-facility location games will be used as subroutines in our mechanisms. Supposing that a set of $n$ agents have single peaked preference over the set of alternative locations $A$, the related results are listed as follows.

\vspace{1.2ex}\noindent\textit{SC-Mechanism~\cite{Feldman2016}.} Given $\mathbf{x}\in\mathcal{R}^n$ and $A$, output $y\in\underset{a\in A}{\arg\min}|a-\mathrm{med}(\mathbf{x})|$, breaking ties in any deterministic way.

\begin{proposition}[\cite{Feldman2016}]\label{single-sc}
SC-Mechanism is group strategyproof and has an approximation ratio of 3 under the sum cost objective.
\end{proposition}

\vspace{1.2ex}\noindent\textit{MC-Mechanism~\cite{Tang2020}.} Given $\mathbf{x}\in\mathcal{R}^n$ and $A$, output $y\in\underset{a\in A}{\arg\min}|a-\mathrm{lt}(\mathbf{x})|$, breaking ties in any deterministic way.

\begin{proposition}[\cite{Tang2020}]\label{single-mc}
MC-Mechanism is group strategyproof and has an approximation ratio of 3 under the maximum cost objective.
\end{proposition}

\subsection{Sum Cost}

\vspace{1.2ex}\noindent\textit{Mechanism 3.}
Given a location and facility preference profile $(\mathbf{x},\mathbf{p})\in\mathcal{R}^n\times\left(2^{\mathcal{F}}\right)^n$, output the facility location profile $\mathbf{y}=(y_1,y_2)$ as follows:
\begin{itemize}
\item[$\bullet$]
if $\left|N_{1,2}\right|>0$, select $(y_{1}, y_{2})\in\underset{(s_{1}, s_{2})\in AP}{\arg \min}\max_{j\in\{1,2\}}\left|s_j-\mathrm{med}\left(\mathbf{x}_{N_{1,2}}\right)\right|$, breaking ties in any deterministic way;
\item[$\bullet$]
if $\left|N_{1,2}\right|=0$ and $\left|N_{1}\right|\geq\left|N_{2}\right|$, select $y_{1}\in\underset{y\in A}{\arg \min }\left|y-\mathrm{med}\left(\mathbf{x}_{N_{1}}\right)\right|$,
and $y_{2}\in\underset{y\in A \backslash\left\{y_{1}\right\}}{\arg \min }\left|y-\mathrm{med}\left(\mathbf{x}_{N_{2}}\right)\right|$ (if $N_2\neq\emptyset$), breaking ties in any deterministic way;
\item[$\bullet$]
if $\left|N_{1,2}\right|=0$ and $\left|N_{1}\right|<\left|N_{2}\right|$, select $y_{2}\in\underset{y\in A}{\arg \min }\left|y-\mathrm{med}\left(\mathbf{x}_{N_{2}}\right)\right|$,
and $y_{1}\in\underset{y\in A \backslash\left\{y_{2}\right\}}{\arg \min }\left|y-\mathrm{med}\left(\mathbf{x}_{N_{1}}\right)\right|$ (if $N_1\neq\emptyset$),
breaking ties in any deterministic way.
\end{itemize}

\begin{theorem}\label{sc-upper-optional}
Mechanism 3 is group strategyproof and has an approximation ratio of at most $2n+1$ under the sum cost objective.
\end{theorem}

\subsection{Maximum Cost}

\vspace{1.2ex}\noindent\textit{Mechanism 4.}
Given a location and facility preference profile $(\mathbf{x},\mathbf{p})\in\mathcal{R}^n\times\left(2^{\mathcal{F}}\right)^n$, output the facility location profile $\mathbf{y}=(y_1,y_2)$ as follows:
\begin{itemize}
\item[$\bullet$]
if $\left|N_{1,2}\right|>0$, select $\left(y_{1}, y_{2}\right)\in\underset{\left(s_{1}, s_{2}\right)\in AP}{\arg \min }\max_{j\in\{1,2\}}\left|s_j-\mathrm{lt}\left(\mathbf{x}_{N_{1,2}}\right)\right|$,
breaking ties in any deterministic way;
\item[$\bullet$]
if $\left|N_{1,2}\right|=0$, select $y_{1}\in\underset{y \in A}{\arg\min}\left|y-\mathrm{lt}\left(\mathbf{x}_{N_{1}}\right)\right|$ (if $N_1\neq\emptyset$),
and $y_{2}\in\underset{y \in A \backslash\left\{y_{1}\right\}}{\arg\min}\left|y-\mathrm{lt}\left(\mathbf{x}_{N_{2}}\right)\right|$ (if $N_2\neq\emptyset$),
breaking ties in any deterministic way.
\end{itemize}

\begin{theorem}\label{mc-upper-optional}
Mechanism 4 is group strategyproof and has an approximation ratio of at most 9 under the maximum cost objective.
\end{theorem}

\section{Conclusion}\label{future}
In this paper, we considered the mechanism design problem for constrained heterogeneous two-facility location games where a set of alternatives are feasible for building facilities and the number of facilities built at each alternative is limited. We studied deterministic mechanisms design without money under the Max-variant cost where the cost of each agent depends on the distance to the further facility. In the compulsory setting where each agent is served by two facilities, we showed that the optimal solution under the sum/maximum cost objective is not strategyproof and proposed a 3-approximate deterministic group strategyproof mechanism which is also the best deterministic strategyproof mechanism for the corresponding social objective. In the optional setting where each agent can be interested in either one of the two facilities or both, we designed a deterministic group strategyproof mechanism with approximation ratio with at most $2n+1$ for the sum cost objective and a deterministic group strategyproof mechanism with approximation ratio with at most 9 for the maximum cost objective.

There are several directions for future research. First, the bounds for approximation ratio of deterministic strategyproof mechanisms in the optional setting do not match yet. Are there more desirable bounds in this setting? Second, randomized mechanism design for constrained heterogeneous facility location games remains an open question. Third, the cost of each agent served by two facilities here is simply the sum of her distances from facilities. How about mechanism design for constrained facility location games in more general settings, such as agents having weighted preference for facilities \cite{Fong2018}? Further, our model can be extended to include more than two facilities or in more general metric spaces.

\subsubsection{Acknowledgements.}
This research was supported in part by the National Natural Science Foundation of China (12171444, 11971447, 11871442), the Natural Science Foundation of Shandong Province of China (ZR2019MA052).

%
%
%
%

\appendix

\section{Missing Proofs}

\subsection{Proof of Lemma \ref{gsp}}

\begin{proof}
Given any $i$, the set of agents $N$ can be divided into $L(i)=\{j\in N\mid x_j<\mathbf{x}_{(i)}\}$, $R(i)=\{j\in N\mid x_j >\mathbf{x}_{(i)}\}$, and $M(i)=\{j\in N\mid x_j=\mathbf{x}_{(i)}\}$.

To show group strategyproofness, we need to prove that for every nonempty $S\subseteq N$ with deviation $\mathbf{x}'_{S}\in\mathcal{R}^{|S|}$, there exists $j\in S$ who cannot benefit from the coalitional deviation. Denote $\mathbf{x}'=(\mathbf{x}'_{S}, \mathbf{x}_{-S})$ and the mechanism by $f$.

\textit{Case 1:} $M(i)\cap S\neq\emptyset$. Then the cost of any agent $j\in M(i)\cap S$ cannot decrease by the deviation since $f(\mathbf{x})$ is her favorite.

\textit{Case 2:} $M(i)\cap S=\emptyset$. If $\mathbf{x}'_{(i)}<\mathbf{x}_{(i)}$, there must exist some agent $j\in R(i)\cap S$ who prefers the peak of $\mathbf{x}_{(i)}$ to that of $\mathbf{x}'_{(i)}$ since $x_j>\mathbf{x}_{(i)}$, which implies that agent $j$ cannot benefit from the deviation. Similarly, if $\mathbf{x}'_{(i)}>\mathbf{x}_{(i)}$, there must exist some agent $j\in L(i)\cap S$ who prefers the peak of $\mathbf{x}_{(i)}$ to that of $\mathbf{x}'_{(i)}$ since $x_j<\mathbf{x}_{(i)}$ and cannot benefit from the deviation.
\end{proof}

\subsection{Proof of Lemma \ref{opt-sc}}

\begin{proof}
Let $OPT_{sc}(\mathbf{x})=(y_1^{\star},y_2^{\star})$ be an optimal solution. Without loss of generality, assume that $y_1^{\star}\le y_2^{\star}$.
Supposing there exists some $a\in A$ such that $y_1^{\star}\le a\le y_2^{\star}$, we only need to show that $sc((y_1^{\star}, a),\mathbf{x})\le sc((y_1^{\star},y_2^{\star}),\mathbf{x})$.

For each agent $i\in N$, $c_i((y_1^{\star}, a),x_i)=\max\{|y_1^{\star}-x_i|,|a-x_i|\}$, $c_i((y_1^{\star}, y_2^{\star}),x_i)=\max\{|y_1^{\star}-x_i|,|y_2^{\star}-x_i|\}$.
If $x_i\le(a+y_2^{\star})/2$, obviously $c_i((y_1^{\star}, a),x_i)\le c_i((y_1^{\star}, y_2^{\star}),x_i)$; otherwise, $c_i((y_1^{\star}, a),x_i)=|y_1^{\star}-x_i|=c_i((y_1^{\star}, y_2^{\star}),x_i)$.

Thus, we have
\begin{eqnarray}
sc((y_1^{\star}, a),\mathbf{x})&=&\sum_{i\in N}c_i((y_1^{\star}, a),x_i)\\
&=&\sum_{i: x_i\le(a+y_2^{\star})/2}c_i((y_1^{\star}, a),x_i)+\sum_{i: x_i>(a+y_2^{\star})/2}c_i((y_1^{\star}, a),x_i)\\
&\le&\sum_{i: x_i\le(a+y_2^{\star})/2}c_i((y_1^{\star}, y_2^{\star}),x_i)+\sum_{i: x_i>(a+y_2^{\star})/2}c_i((y_1^{\star}, y_2^{\star}),x_i)\\
&=&sc((y_1^{\star},y_2^{\star}),\mathbf{x})
\end{eqnarray}
\end{proof}

\subsection{Proof of Theorem \ref{sc-lower}}

\begin{proof}
Suppose $f$ is a deterministic strategyproof mechanism with approximation ratio of $3-\delta$ for some $\delta>0$.

Consider an instance $I(\mathbf{x},A)$ with $\mathbf{x}=(-\varepsilon, \varepsilon)$ and $A=\{-1,-1,1,1\}$, where $\varepsilon>0$ is sufficiently small.
$f(\mathbf{x})$ can be $(-1,-1)$, $(1,1)$, $(-1,1)$, or $(1,-1)$ and assume w.l.o.g. that $f(\mathbf{x})=(1,1)$ or $(-1,1)$. Then the cost of agent 1 is $c_1(f(\mathbf{x}),x_1)=1+\varepsilon$.

For another instance $I(\mathbf{x}',A)$ with $\mathbf{x}^{\prime}=(-1, \varepsilon)$, it holds that $OPT_{sc}(\mathbf{x}')=(-1,-1)$ and $sc(OPT,\mathbf{x}')=1+\varepsilon$.
If $f(\mathbf{x}')=(1,1)$, $(-1,1)$, or $(1,-1)$, then $sc(f,\mathbf{x}')\ge 3-\varepsilon$. This implies that
\begin{equation}
\frac{sc(f,\mathbf{x}')}{sc(OPT,\mathbf{x}')}\ge\frac{3-\varepsilon}{1+\varepsilon}>3-\delta
\end{equation}
for sufficiently small $\varepsilon>0$, which is a contradiction. Thus, $f(\mathbf{x}')=(-1,-1)$.

Note that $c_1(f(\mathbf{x}'),x_1)=1-\varepsilon$.
This indicates that agent 1 can decrease her cost by misreporting her location as $x_{1}'=-1$, which contradicts $f$'s strategyproofness.
\end{proof}

\subsection{Proof of Theorem \ref{sc-upper}}

\begin{proof}
By Lemma \ref{gsp}, \textit{Mechanism 1} is group strategyproof. We now turn to its approximation ratio.

Given a location profile $\mathbf{x}\in\mathcal{R}^n$, let $OPT_{sc}(\mathbf{x})=(y_1^{\star},y_2^{\star})\in AP$ be an optimal solution. Denote \textit{Mechanism 1} by $f$ and $f(\mathbf{x})=(y_1, y_2)$.

Considering that both $f(\mathbf{x})$ and $OPT_{sc}(\mathbf{x})$ are adjacent location pairs in $AP$, assume w.l.o.g. that $(y_1^{\star},y_2^{\star})$ is on the right of $(y_1, y_2)$.

Let $y_2'\in A$ be the location adjacent to the right of $y_2$ and $y'=(y_1+y_2')/2$ be the right border of the zone of $(y_1,y_2)$. We first give two claims, then compare $sc(f,\mathbf{x})$ with $sc(OPT,\mathbf{x})$.

\vspace{1.1ex}\noindent\textbf{Claim 1.} $\left|\left\{i \in N \mid x_{i} \leq y'\right\}\right| \geq\left|\left\{i \in N \mid x_{i}>y'\right\}\right|$, since $\mathrm{med}(\mathbf{x})\le y'$.

\vspace{1.1ex}\noindent\textbf{Claim 2.} For any agent $i$ with $x_i\le y'$, it holds that $c_i(f(\mathbf{x}),x_i)\le c_i(OPT_{sc}(\mathbf{x}),x_i)$, since the peak of agent $i$ in $AP$ is $(y_1, y_2)$ or to the left.

The sum cost of \textit{Mechanism 1} is
\begin{eqnarray}
sc(f,\mathbf{x})
&=&\sum_{i\in N} c_{i}\left(\left(y_{1}, y_{2}\right),x_{i})\right)=\sum_{i\in N} \max _{j \in\{1,2\}}\left|x_{i}-y_{j}\right| \\
&=&\sum_{x_{i} \leq y'} \max _{j \in\{1,2\}}\left|x_{i}-y_{j}\right|+\sum_{x_{i}>y'} \max _{j \in\{1,2\}}\left|x_{i}-y_{j}\right|,
\end{eqnarray}
where the first term is denoted by $\alpha$ and the second by $\beta$.

The optimal sum cost is
\begin{eqnarray}
sc(OPT,\mathbf{x})
&=&\sum_{i\in N} c_{i}\left(\left({y}_{1}^{\star}, {y}_{2}^{\star}\right),x_{i})\right)=\sum_{i\in N} \max _{j \in\{1,2\}}\left|x_{i}-{y}_{j}^{\star}\right| \\
&=& \sum_{x_{i} \leq y'} \max _{j \in\{1,2\}}\left|x_{i}-{y}_{j}^{\star}\right|+\sum_{x_{i}>y'} \max _{j \in\{1,2\}}\left|x_{i}-{y}_{j}^{\star}\right|,
\end{eqnarray}
where the first term is denoted by $\gamma$ and the second by $\delta$.

Note that
\begin{eqnarray}
\beta &=&\sum_{x_{i}>y'} \max _{j \in\{1,2\}}\left|x_{i}-y_{j}\right|\leq \sum_{x_{i}>y'} \max _{j \in\{1,2\}}\left\{\left|x_{i}-{y}^{\star}\right|+\left|{y}^{\star}-y_{j}\right|\right\} \\
&\leq &\sum_{x_{i}>y'}\left|x_{i}-{y}^{\star}\right|+\sum_{x_{i}>y'} \max _{j \in\{1,2\}}\left|{y}^{\star}-y_{j}\right| \\
&\leq& \sum_{x_{i}>y'}\left|x_{i}-{y}^{\star}\right|+\sum_{x_{i} \leq y'} \max _{j \in\{1,2\}}\left|{y}^{\star}-y_{j}\right| \\
&\leq &\sum_{x_{i}>y'}\left|x_{i}-{y}^{\star}\right|+\sum_{x_{i} \leq y'} \max _{j \in\{1,2\}}\left\{\left|{y}^{\star}-x_{i}\right|+\left|x_{i}-y_{j}\right|\right\} \\
&\leq &\sum_{i\in N}\left|x_{i}-{y}^{\star}\right|+\sum_{x_{i} \leq y'} \max _{j \in\{1,2\}}\left|x_{i}-y_{j}\right| \\
&\leq &\sum_{i\in N}\max _{j \in\{1,2\}}\left|x_{i}-{y}_{j}^{\star}\right|+\sum_{x_{i} \leq y'} \max _{j \in\{1,2\}}\left|x_{i}-y_{j}\right|=\gamma+\delta+\alpha.
\end{eqnarray}

Here, the third inequality holds by Claim 1. Besides, we have $\alpha \leq \gamma$ by Claim 2. Thus,
\begin{equation}
\frac{sc(f,\mathbf{x})}{sc(O P T,\mathbf{x})}=\frac{\alpha+\beta}{\gamma+\delta} \leq \frac{\alpha+\gamma+\delta+\alpha}{\gamma+\delta} \leq \frac{3 \gamma+\delta}{\gamma+\delta} \leq 3
\end{equation}
Combining with Theorem \ref{sc-lower}, the approximation ratio of \textit{Mechanism 1} is 3.
\end{proof}

\subsection{Proof of Lemma \ref{opt-mc}}

\begin{proof}
Let $\mathbf{a}=(a_k,a_{k+1})$ be the peak of $\mathrm{cen}(\mathbf{x})$ in $AP$. If there exists $s(\in A)<a_k$, then
\begin{equation}\label{left}
(s+a_{k+1})/2\le(a_{k-1}+a_{k+1})/2\le\mathrm{cen}(\mathbf{x}).
\end{equation}
If there exists $s(\in A)>a_{k+1}$, then
\begin{equation}\label{right}
(a_k+s)/2\ge(a_{k}+a_{k+2})/2\ge\mathrm{cen}(\mathbf{x}).
\end{equation}

Let $\mathbf{y}=(y_1,y_2)$ be any feasible solution that is different from $(a_k,a_{k+1})$. Assume w.l.o.g. that $y_1\le y_2$, then either $y_1<a_k$ or $a_{k+1}<y_2$. By symmetry, we only need to compare $mc(\mathbf{y},\mathbf{x})$ with $mc(\mathbf{a},\mathbf{x})$ through the following two cases.

\textit{Case 1:} $a_k\le a_{k+1}\le\mathrm{cen}(\mathbf{x})$. In this case, $mc(\mathbf{a},\mathbf{x})=\mathrm{rt}(\mathbf{x})-a_k$.
If $y_1<a_k$, then
 \begin{equation}
 mc(\mathbf{y},\mathbf{x})\ge \mathrm{rt}(\mathbf{x})-y_1>\mathrm{rt}(\mathbf{x})-a_k=mc(\mathbf{a},\mathbf{x}).
 \end{equation}
If $a_{k+1}<y_2$, then $y_2-\mathrm{cen}(\mathbf{x})\ge \mathrm{cen}(\mathbf{x})-a_k$ by Eq. (\ref{right}). Thus, we have
\begin{eqnarray}
mc(\mathbf{y},\mathbf{x})&\ge& y_2-\mathrm{lt}(\mathbf{x})=y_2-\mathrm{cen}(\mathbf{x})+\mathrm{cen}(\mathbf{x})-\mathrm{lt}(\mathbf{x})\\
&\ge&\mathrm{cen}(\mathbf{x})-a_k+\mathrm{rt}(\mathbf{x})-\mathrm{cen}(\mathbf{x})=mc(\mathbf{a},\mathbf{x}).
\end{eqnarray}

\textit{Case 2:} $a_k\le\mathrm{cen}(\mathbf{x})<a_{k+1}$. In this case,
$mc(\mathbf{a},\mathbf{x})=\max\{\mathrm{rt}(\mathbf{x})-a_k, a_{k+1}-\mathrm{lt}(\mathbf{x})\}$.
If $y_1<a_k$, then $\mathrm{rt}(\mathbf{x})-y_1>\mathrm{rt}(\mathbf{x})-a_k$ and by Eq. (\ref{left}), it holds that
\begin{eqnarray}
\mathrm{rt}(\mathbf{x})-y_1&=&\mathrm{rt}(\mathbf{x})-\mathrm{cen}(\mathbf{x})+\mathrm{cen}(\mathbf{x})-y_1\\
&\ge&\mathrm{cen}(\mathbf{x})-\mathrm{lt}(\mathbf{x})+a_{k+1}-\mathrm{cen}(\mathbf{x})
=a_{k+1}-\mathrm{lt}(\mathbf{x}).
\end{eqnarray}
Thus, we have $mc(\mathbf{y},\mathbf{x})\ge \mathrm{rt}(\mathbf{x})-y_1\ge mc(\mathbf{a},\mathbf{x})$.
Similarly if $a_{k+1}<y_2$, then $y_2-\mathrm{lt}(\mathbf{x})>a_{k+1}-\mathrm{lt}(\mathbf{x})$ and
$y_2-\mathrm{lt}(\mathbf{x})=y_2-\mathrm{cen}(\mathbf{x})+\mathrm{cen}(\mathbf{x})-\mathrm{lt}(\mathbf{x})
\ge\mathrm{cen}(\mathbf{x})-a_k+\mathrm{rt}(\mathbf{x})-\mathrm{cen}(\mathbf{x})=\mathrm{rt}(\mathbf{x})-a_k$. Thus, we have $mc(\mathbf{y},\mathbf{x})\ge y_2-\mathrm{lt}(\mathbf{x})\ge mc(\mathbf{a},\mathbf{x})$.
\end{proof}

\subsection{Proof of Theorem \ref{mc-lower}}

\begin{proof}
Suppose $f$ is a deterministic strategyproof mechanism with approximation ratio of $3-\delta$ for some $\delta>0$.

Consider an instance $I(\mathbf{x},A)$ with $\mathbf{x}=(-\varepsilon, \varepsilon)$ and $A=\{-1,-1,1,1\}$, where $\varepsilon>0$ is sufficiently small.
$f(\mathbf{x})$ can be $(-1,-1)$, $(1,1)$, $(-1,1)$, or $(1,-1)$ and assume w.l.o.g. that $f(\mathbf{x})=(1,1)$ or $(-1,1)$. Then the cost of agent 1 is $c_1(f(\mathbf{x}),x_1)=1+\varepsilon$.

For another instance $I(\mathbf{x}',A)$ with $\mathbf{x}^{\prime}=(-2-\varepsilon, \varepsilon)$, it holds that $OPT_{mc}(\mathbf{x}')=(-1,-1)$ and $mc(OPT,\mathbf{x}')=1+\varepsilon$.
If $f(\mathbf{x}')=(1,1)$, $(-1,1)$, or $(1,-1)$, then $mc(f,\mathbf{x}')=3+\varepsilon$. This implies that
\begin{equation}
\frac{mc(f,\mathbf{x}')}{mc(OPT,\mathbf{x}')}=\frac{3+\varepsilon}{1+\varepsilon}>3-\delta
\end{equation}
for sufficiently small $\varepsilon>0$, which is a contradiction. Thus, $f(\mathbf{x}')=(-1,-1)$.

Considering that $c_1(f(\mathbf{x}'),x_1)=1-\varepsilon$, agent 1 can decrease her cost by misreporting her location as $x_{1}'=-2-\varepsilon$, which contradicts $f$'s strategyproofness.
\end{proof}

\subsection{Proof of Theorem \ref{mc-upper}}

\begin{proof}

By Lemma \ref{gsp}, \textit{Mechanism 2} is group strategyproof. We now turn to its approximation ratio.

Given a location profile $\mathbf{x}\in\mathcal{R}^n$, let $OPT_{mc}(\mathbf{x})=(y_1^{\star},y_2^{\star})$ be the peak of $\mathrm{cen}(\mathbf{x})$ in $AP$ which is also an optimal solution. Denote \textit{Mechanism 2} by $f$ and $f(\mathbf{x})=(y_1, y_2)$.
Assume without loss of generality that $\mathrm{rt}(\mathbf{x})-\mathrm{lt}(\mathbf{x})=1$.

It is easy to see that $mc(OPT,\mathbf{x}) \geq \frac{1}{2}$, and
\begin{equation}
mc(OPT,\mathbf{x}) \geq \max_{j\in\{1,2\}}\left|\mathrm{lt}(\mathbf{x})-y_{j}^{\star}\right| \ge\max _{j \in\{1,2\}}\left|\mathrm{lt}(\mathbf{x})-y_{j}\right|.
\end{equation}

We compare $mc(f,\mathbf{x})$ with $mc(OPT,\mathbf{x})$ through the following analysis.

\textit{Case 1:} $y_{1} \leq y_{2} \leq \mathrm{lt}(\mathbf{x}) \leq \mathrm{rt}(\mathbf{x})$, or $y_{1} \leq \mathrm{lt}(\mathbf{x}) \leq y_{2} \leq \mathrm{rt}(\mathbf{x})$.
\begin{equation}
mc(f,\mathbf{x})=\left|\mathrm{rt}(\mathbf{x})-y_{1}\right|=1+\left|\mathrm{lt}(\mathbf{x})-y_{1}\right| \leq 3 mc(OPT,\mathbf{x}).
\end{equation}

\textit{Case 2:} $\mathrm{lt}(\mathbf{x}) \leq y_{1} \leq y_{2} \leq \mathrm{rt}(\mathbf{x})$.
\begin{equation}
m c(f,\mathbf{x}) \leq 1 \leq 2 mc(OPT,\mathbf{x}).
\end{equation}

\textit{Case 3:} $\mathrm{lt}(\mathbf{x}) \leq y_{1} \leq \mathrm{rt}(\mathbf{x}) \leq y_{2}$, or $\mathrm{lt}(\mathbf{x}) \leq \mathrm{rt}(\mathbf{x}) \leq y_{1} \leq y_{2}$.

In this case, the right border of the zone of $(y_1,y_2)$ is no less than $(y_1+y_2)/2\ge \mathrm{cen}(\mathbf{x})\ge\mathrm{lt}(\mathbf{x})$. Combining with the fact that $\mathrm{lt}(\mathbf{x})$ lies in the zone of $(y_1,y_2)$, it holds that $\mathrm{cen}(\mathbf{x})$ also lies in the zone of $(y_1,y_2)$. This implies that $f(\mathbf{x})=OPT_{mc}(\mathbf{x})$. Thus, we have
\begin{equation}
mc(f,\mathbf{x})=mc(OPT,\mathbf{x}).
\end{equation}

\textit{Case 4:} $y_{1} \leq \mathrm{lt}(\mathbf{x}) \leq \mathrm{rt}(\mathbf{x}) \leq y_{2}$.
Note that
\begin{equation}
|\mathrm{lt}(\mathbf{x})-y_{2}|\le \max_{j\in\{1,2\}}|y_j-\mathrm{lt}(\mathbf{x})|\le\max_{j\in\{1,2\}}|y_j^{\star}-\mathrm{lt}(\mathbf{x})|\le mc(OPT,\mathbf{x}),
\end{equation}
and
\begin{equation}
\left|\mathrm{rt}(\mathbf{x})-y_{1}\right|=1+\left|\mathrm{lt}(\mathbf{x})-y_{1}\right|\le 1+mc(OPT,\mathbf{x})\le 3mc(OPT,\mathbf{x}).
\end{equation}
Thus, it holds that
\begin{equation}
mc(f,\mathbf{x})=\max \left\{\left|\mathrm{lt}(\mathbf{x})-y_{2}\right|,\left|\mathrm{rt}(\mathbf{x})-y_{1}\right|\right\}|\le 3mc(OPT,\mathbf{x}).
\end{equation}

Above all, $mc(f,\mathbf{x}) \leq 3 mc(OPT,\mathbf{x})$. Combining with Theorem \ref{mc-lower}, \textit{Mechanism 2} has an approximation ratio of 3.
\end{proof}

\subsection{Proof of Theorem \ref{sc-upper-optional}}

\begin{proof}
\textbf{Group strategyproofness.} Given $(\mathbf{x},\mathbf{p})\in\mathcal{R}^n\times\left(2^{\mathcal{F}}\right)^n$, \textit{Mechanism 3} outputs the facility location profile according to the public information $\mathbf{p}$. To show group strategyproofness, we need to prove that for every nonempty $S\subseteq N$ with deviation $\mathbf{x}'_{S}\in\mathcal{R}^{|S|}$, there exists $j\in S$ who cannot benefit from the coalitional deviation. Denote $\mathbf{x}'=(\mathbf{x}'_{S}, \mathbf{x}_{-S})$, \textit{Mechanism 3} by $f$, \textit{Mechanism 1} by $f_1$, and \textit{SC-Mechanism} by $f_2$.

\textit{Case 1:} $|N_{1,2}|>0$, then $f(\mathbf{x},\mathbf{p})=f_1(\mathbf{x}_{N_{1,2}})$ and $f(\mathbf{x}',\mathbf{p})=f_1(\mathbf{x}'_{N_{1,2}\cap S},\mathbf{x}_{N_{1,2}\backslash S})$. If $N_{1,2}\cap S\neq\emptyset$, any agent in $N_{1,2}\cap S$ cannot benefit from the deviation $\mathbf{x}'_{N_{1,2}\cap S}$ by $f_1$'s group strategyproofness. If $N_{1,2}\cap S=\emptyset$, $f(\mathbf{x}',\mathbf{p})=f_3(\mathbf{x}_{N_{1,2}})$, which implies that any agent in $S\subseteq N_1\cup N_2$ cannot benefit from the deviation.

\textit{Case 2:} $|N_{1,2}|=0$ and $|N_{1}|\geq|N_{2}|$. It holds that $f(\mathbf{x},\mathbf{p})=(f_2(\mathbf{x}_{N_1}),f_2(\mathbf{x}_{N_2}))$
and $f(\mathbf{x}',\mathbf{p})=(f_2(\mathbf{x}'_{N_1\cap S},\mathbf{x}_{N_1\backslash S}),f_2(\mathbf{x}'_{N_2\cap S},\mathbf{x}_{N_2\backslash S}))$,
with $f_2(\mathbf{x}_{N_2})\in A\backslash f_2(\mathbf{x}_{N_1})$
and $f_2(\mathbf{x}'_{N_2\cap S},\mathbf{x}_{N_2\backslash S})\in A\backslash f_2(\mathbf{x}'_{N_1\cap S},\mathbf{x}_{N_1\backslash S})$.
If $N_1\cap S\neq\emptyset$, any agent in $N_1\cap S$ cannot benefit from the deviation $\mathbf{x}'_{N_1\cap S}$ by $f_2$'s group strategyproofness.
If $N_1\cap S=\emptyset$,
$f(\mathbf{x}',\mathbf{p})=(f_2(\mathbf{x}_{N_1}),f_2(\mathbf{x}'_{N_2\cap S},\mathbf{x}_{N_2\backslash S}))$
with $f_2(\mathbf{x}'_{N_2\cap S},\mathbf{x}_{N_2\backslash S})\in A\backslash f_2(\mathbf{x}_{N_1})$.
Still by $f_2$'s group strategyproofness, any agent in $N_2\cap S$ cannot benefit from the deviation $\mathbf{x}'_{N_2\cap S}$.

\textit{Case 3:} $|N_{1,2}|=0$ and $|N_{1}|<|N_{2}|$. This case is similar to \textit{Case 2}.

\textbf{Approximation ratio.}
Given $(\mathbf{x},\mathbf{p})\in\mathcal{R}^n\times\left(2^{\mathcal{F}}\right)^n$, let $OPT_{sc}(\mathbf{x},\mathbf{p})=\mathbf{y}^{\star}=(y_1^{\star},y_2^{\star})$ be an optimal solution and $f(\mathbf{x},\mathbf{p})=\mathbf{y}=(y_1, y_2)$. We now compare $sc(f,(\mathbf{x},\mathbf{p}))$ with $sc(OPT,(\mathbf{x},\mathbf{p}))$.

\textit{Case 1:} If $\left|N_{1,2}\right|>0$, the output of \textit{Mechanism 3} on $I(\mathbf{x},\mathbf{p},A)$ equals to that of \textit{Mechanism 1} on $I\left(\mathbf{x}_{N_{1,2}},\mathbf{p}_{N_{1,2}},A\right)$. Denote the optimal solution on $I\left(\mathbf{x}_{N_{1,2}},\mathbf{p}_{N_{1,2}},A\right)$ as $\mathbf{y}^{opt}$.

By Theorem \ref{sc-upper}, it holds that
\begin{equation}
\sum_{i \in N_{1,2}} c_{i}\left(\mathbf{y},\left(x_{i}, p_{i}\right)\right) \leq 3 \sum_{i \in N_{1,2}} c_{i}\left(\mathbf{y}^{opt},\left(x_{i}, p_{i}\right)\right) \leq 3 \sum_{i \in N_{1,2}} c_{i}\left(\mathbf{y}^{\star},\left(x_{i}, p_{i}\right)\right).
\end{equation}

Thus, we have
\begin{eqnarray}
s c(f,(\mathbf{x}, \mathbf{p})) &=&\sum_{i \in N_{1} \cup N_{2} \cup N_{1,2}} c_{i}\left(\mathbf{y},\left(x_{i}, p_{i}\right)\right) \\
& \leq &\sum_{i \in N_{1}}\left|x_{i}-y_{1}\right|+\sum_{i \in N_{2}}\left|x_{i}-y_{2}\right|+3 \sum_{i \in N_{1,2}} c_{i}\left(\mathbf{y}^{\star},\left(x_{i}, p_{i}\right)\right) \\
& \leq &\sum_{i \in N_{1}}\left|x_{i}-y_{1}^{\star}\right|+\sum_{i \in N_{2}}\left|x_{i}-y_{2}^{\star}\right|+3 \sum_{i \in N_{1,2}} c_{i}\left(\mathbf{y}^{\star},\left(x_{i}, p_{i}\right)\right)\\
&&+\left|N_{1}\right| \cdot\left|y_{1}-y_{1}^{\star}\right|+\left|N_{2}\right| \cdot\left|y_{2}-y_{2}^{\star}\right| \\
& \leq &3 sc(OPT,(\mathbf{x}, \mathbf{p}))+\left|N_{1} \cup N_{2}\right| \cdot 2 sc(OPT,(\mathbf{x}, \mathbf{p})) \\
& \leq&(2n+1) sc(OPT,(\mathbf{x}, \mathbf{p})).
\end{eqnarray}
Here, the above third inequality holds because for $j=1,2$,
\begin{eqnarray}
\left|y_{j}-y_{j}^{\star}\right| &\leq&\left|y_{j}-\mathrm{med}\left(\mathbf{x}_{N_{1,2}}\right)\right|+
\left|\mathrm{med}\left(\mathbf{x}_{N_{1,2}}\right)-y_{j}^{\star}\right|\\
&\le&\max_{k\in\{1,2\}}\left|y_k-\mathrm{med}\left(\mathbf{x}_{N_{1,2}}\right)\right|+\left|\mathrm{med}\left(\mathbf{x}_{N_{1,2}}\right)-y_{j}^{\star}\right|\\
&\le&\max_{k\in\{1,2\}}\left|y_k^{\star}-\mathrm{med}\left(\mathbf{x}_{N_{1,2}}\right)\right|+\left|\mathrm{med}\left(\mathbf{x}_{N_{1,2}}\right)-y_{j}^{\star}\right|\\
&\leq& 2 sc(OPT,(\mathbf{x}, \mathbf{p})).
\end{eqnarray}

\textit{Case 2:} If $\left|N_{1,2}\right|=0$ and $\left|N_{1}\right| \geq\left|N_{2}\right|$. Without loss of generality, assume that $N_2\neq\emptyset$. $y_1$ equals to the output of \textit{SC-Mechanism} on instance $I_1=I\left(\mathbf{x}_{N_1},\mathbf{p}_{N_1},A\right)$, and $y_2$ equals to the output of \textit{SC-Mechanism} on instance $I_2=I\left(\mathbf{x}_{N_2},\mathbf{p}_{N_2},A\backslash\{y_1\}\right)$. Denote by $y_{1}^{opt}$ the optimal solution on instance $I_1$ and $y_{2}^{opt}$ the optimal solution on instance $I_2$.

For $k=1,2$, let $sc(y,I_k)=\sum_{i\in N_k}|x_i-y|$, then
\begin{eqnarray}
sc(OPT,(\mathbf{x}, \mathbf{p}))&=&\sum_{i\in N_{1}}\left|x_{i}-{y}_{1}^{\star}\right|+\sum_{i \in N_{2}}\left|x_{i}-{y}_{2}^{\star}\right|=sc\left({y}_{1}^{\star}, I_{1}\right)+sc\left({y}_{2}^{\star}, I_{2}\right) \\
sc(f,(\mathbf{x}, \mathbf{p})) &=&\sum_{i \in N_{1}}\left|x_{i}-y_{1}\right|+\sum_{i \in N_{2}}\left|x_{i}-y_{2}\right|=sc\left(y_{1}, I_{1}\right)+sc\left(y_{2}, I_{2}\right).
\end{eqnarray}

For $I_1$, by Proposition \ref{single-sc}, it holds that
\begin{equation}\label{I1-sc}
sc\left(y_{1}, I_{1}\right) \leq 3 sc\left(y_{1}^{opt}, I_{1}\right)\le 3 sc\left({y}_{1}^{\star}, I_{1}\right)
\end{equation}

For $I_2$, we consider the following two cases.

\textit{Case 2.1:} If ${y}_{2}^{\star} \in A \backslash\left\{y_{1}\right\}$, by Proposition \ref{single-sc}, it holds that
\begin{equation}
sc\left(y_{2}, I_{2}\right) \leq 3 sc\left(y_{2}^{opt}, I_{2}\right)\le 3 sc\left({y}_{2}^{\star}, I_{2}\right)
\end{equation}

\textit{Case 2.2:} ${y}_{2}^{\star} \notin A \backslash\left\{y_{1}\right\}$, then $y_{1}={y}_{2}^{\star}$ and ${y}_{1}^{\star}\in A \backslash\left\{y_{1}\right\}$.
On the one hand, by Proposition \ref{single-sc}, we have
\begin{equation}\label{I2-sc}
sc\left(y_{2}, I_{2}\right)\le 3 sc\left(y_{2}^{opt}, I_{2}\right)\le 3 sc\left({y}_{1}^{\star}, I_{2}\right).
\end{equation}
On the other hand,
\begin{eqnarray}
sc\left({y}_{1}^{\star}, I_{2}\right)&=& \sum_{i \in N_{2}}\left|x_{i}-{y}_{1}^{\star}\right| \leq \sum_{i \in N_{2}}\left|x_{i}-{y}_{2}^{\star}\right|+\sum_{i \in N_{1}}\left|{y}_{2}^{\star}-{y}_{1}^{\star}\right| \\
& \leq& \sum_{i \in N_{2}}\left|x_{i}-{y}_{2}^{\star}\right|+\sum_{i \in N_{1}}\left|y_{1}-x_{i}\right|+\sum_{i \in N_{1}}\left|x_{i}-{y}_{1}^{\star}\right| \\
&=&sc\left({y}_{2}^{\star}, I_{2}\right)+sc\left({y}_1, I_{1}\right)+sc\left({y}_{1}^{\star}, I_{1}\right)\\
& \leq& sc\left({y}_{2}^{\star}, I_{2}\right)+4 sc\left({y}_{1}^{\star}, I_{1}\right),\label{I12-sc}
\end{eqnarray}
where the first inequality holds because $\left|N_{1}\right| \geq\left|N_{2}\right|$ and the third holds by Eq. (\ref{I1-sc}).

Combining Eq. (\ref{I2-sc}) and Eq. (\ref{I12-sc}), we have
\begin{equation}\label{I221-sc}
sc\left(y_{2}, I_{2}\right) \leq 3 sc\left({y}_{2}^{\star}, I_{2}\right)+12 sc\left({y}_{1}^{\star}, I_{1}\right)
\end{equation}

Thus, by Eq. (\ref{I1-sc}) and Eq. (\ref{I221-sc}), it holds that
\begin{eqnarray}
s c(f,(\mathbf{x}, \mathbf{p}))&=&sc\left(y_{1}, I_{1}\right)+sc\left(y_{2}, I_{2}\right) \\
& \leq& 3 sc\left({y}_{1}^{\star}, I_{1}\right)+3 sc\left({y}_{2}^{\star}, I_{2}\right)+12 sc\left({y}_{1}^{\star}, I_{1}\right) \\
&\leq& 15 sc(OPT,(\mathbf{x}, \mathbf{p}))
\end{eqnarray}

\textit{Case 3:} $\left|N_{1,2}\right|=0$ and $\left|N_{1}\right|<\left|N_{2}\right|$. This case is similar to \textit{Case 2}.

Above all, \textit{Mechanism 3} has an approximation ratio of at most $2n+1$.
\end{proof}

\subsection{Proof of Theorem \ref{mc-upper-optional}}

\begin{proof}
The proof of \textit{Mechanism 4}'s group strategyproofness is similar to that of \textit{Mechanism 3}'s, which is omitted here. Now we focus on the approximation ratio of \textit{Mechanism 4}.

Denote \textit{Mechanism 4} by $f$.
Given $(\mathbf{x},\mathbf{p})\in\mathcal{R}^n\times\left(2^{\mathcal{F}}\right)^n$, let $OPT_{mc}(\mathbf{x},\mathbf{p})=\mathbf{y}^{\star}=(y_1^{\star},y_2^{\star})$ be an optimal solution and $f(\mathbf{x},\mathbf{p})=\mathbf{y}=(y_1, y_2)$. We now compare $mc(f,(\mathbf{x},\mathbf{p}))$ with $mc(OPT,(\mathbf{x},\mathbf{p}))$.

\textit{Case 1:} If $\left|N_{1,2}\right|>0$, the output of \textit{Mechanism 4} on $I(\mathbf{x},\mathbf{p},A)$ equals to that of \textit{Mechanism 2} on $I\left(\mathbf{x}_{N_{1,2}},\mathbf{p}_{N_{1,2}},A\right)$. Denote by $\mathbf{y}^{opt}=(y_1^{opt},y_2^{opt})$ the optimal solution on $I\left(\mathbf{x}_{N_{1,2}},\mathbf{p}_{N_{1,2}},A\right)$.

By Theorem \ref{mc-upper}, it holds that
\begin{equation}
\max _{i \in N_{1,2}} c_{i}\left(\mathbf{y},\left(x_{i}, p_{i}\right)\right) \leq 3 \max _{i \in N_{1,2}} c_{i}\left(\mathbf{y}^{opt},\left(x_{i}, p_{i}\right)\right)\leq 3 \max _{i \in N_{1,2}} c_{i}\left(\mathbf{y}^{\star},\left(x_{i}, p_{i}\right)\right).
\end{equation}

Thus, we have
\begin{eqnarray}
&&mc(f,(\mathbf{x}, \mathbf{p}))\\
&=&\max_{i \in N_{1} \cup N_{2} \cup N_{1,2}}\left\{c_{i}\left(\mathbf{y},\left(x_{i}, p_{i}\right)\right)\right\}\\
&=&\max \left\{ \max_{i \in N_{1}}\left\{\left|y_{1}-x_{i}\right|\right\}, \max_{i \in N_{2}}\left\{| y_{2}-x_{i}|\right\}, \max_{i \in N_{1,2}}\left\{c_{i}(\mathbf{y},(x_{i}, p_{i}))\right\}\right\}\\
&\leq& \max \left\{\max _{i \in N_{1}}\left\{\left|y_{1}^{\star}-x_{i}\right|+\left|y_{1}-y_{1}^{\star}\right|\right\}, \max _{i \in N_{2}}\left\{\left|y_{2}^{\star}-x_{i}\right|\right.\right.\\
&&\left.\left.+\left|y_{2}-y_{2}^{\star}\right|\right\}, 3 \max _{i \in N_{1,2}} c_{i}\left(\mathbf{y}^{\star},\left(x_{i}, p_{i}\right)\right)\right\}\\
&\leq& \max \left\{\max _{i \in N_{1}}\left\{\left|y_{1}^{\star}-x_{i}\right|+2 mc(OPT,(\mathbf{x}, \mathbf{p}))\right\}, \max _{i \in N_{2}}\left\{\left|y_{2}^{\star}-x_{i}\right|\right.\right.\\
&&\left.+2 mc(O P T,(\mathbf{x}, \mathbf{p}))\}, 3 \max _{i \in N_{1,2}} c_{i}\left(\mathbf{y}^{\star},\left(x_{i}, p_{i}\right)\right)\right\}\\
 &\leq& 3 mc(O P T,(\mathbf{x}, \mathbf{p})).
\end{eqnarray}

Here, the above second inequality holds because for $j=1,2$,
\begin{eqnarray}
\left|y_{j}-y_{j}^{\star}\right| &\leq&\left|y_{j}-\mathrm{lt}\left(\mathbf{x}_{N_{1,2}}\right)\right|+
\left|\mathrm{lt}\left(\mathbf{x}_{N_{1,2}}\right)-y_{j}^{\star}\right|\\
&\le&\max_{k\in\{1,2\}}\left|y_k-\mathrm{lt}\left(\mathbf{x}_{N_{1,2}}\right)\right|+\left|\mathrm{lt}\left(\mathbf{x}_{N_{1,2}}\right)-y_{j}^{\star}\right|\\
&\le&\max_{k\in\{1,2\}}\left|y_k^{\star}-\mathrm{lt}\left(\mathbf{x}_{N_{1,2}}\right)\right|+\left|\mathrm{lt}\left(\mathbf{x}_{N_{1,2}}\right)-y_{j}^{\star}\right|\\
&\leq& 2 mc(OPT,(\mathbf{x}, \mathbf{p})).
\end{eqnarray}

\textit{Case 2:} $\left|N_{1,2}\right|=0$. Assume w.l.o.g. that $N_1\neq\emptyset, N_2\neq\emptyset$.
$y_1$ equals to the output of \textit{MC-Mechanism} on instance $I_1=I\left(\mathbf{x}_{N_1},\mathbf{p}_{N_1},A\right)$, and $y_2$ equals to the output of \textit{MC-Mechanism} on instance $I_2=I\left(\mathbf{x}_{N_2},\mathbf{p}_{N_2},A\backslash\{y_1\}\right)$.
Denote by $y_{1}^{opt}$ the optimal solution on instance $I_1$ and $y_{2}^{opt}$ the optimal solution on instance $I_2$.

For $k=1,2$, let $mc(y,I_k)=\max _{i \in N_{k}}\left|x_{i}-y\right|$, then
\begin{eqnarray}
mc(O P T,(\mathbf{x}, \mathbf{p}))&=&\max \left\{\max _{i \in N_{1}}\left|x_{i}-y_{1}^{\star}\right|, \max _{i \in N_{2}}\left|x_{i}-y_{2}^{\star}\right|\right\} \\
&=&\max \left\{mc\left({y}_{1}^{\star}, I_{1}\right), mc\left({y}_{2}^{\star}, I_{2}\right)\right\} \\
mc(f,(\mathbf{x}, \mathbf{p}))&=&\max \left\{mc\left(y_{1}, I_{1}\right), mc\left(y_{2}, I_{2}\right)\right\}
\end{eqnarray}

For $I_1$, by Proposition \ref{single-mc}, it holds that
\begin{equation}\label{I1-mc}
mc\left(y_{1}, I_{1}\right) \leq 3mc\left(y_{1}^{opt}, I_{1}\right)\le 3 mc\left({y}_{1}^{\star}, I_{1}\right)
\end{equation}

For $I_2$, we consider the following two cases.

\textit{Case 2.1:} If ${y}_{2}^{\star} \in A \backslash\left\{y_{1}\right\}$, by Proposition \ref{single-mc}, it holds that
\begin{eqnarray}
mc\left(y_{2}, I_{2}\right) \leq 3 mc\left(y_{2}^{opt}, I_{2}\right)\le 3 mc\left({y}_{2}^{\star}, I_{2}\right)
\end{eqnarray}

\textit{Case 2.2:} ${y}_{2}^{\star} \notin A \backslash\left\{y_{1}\right\}$, then $y_{1}={y}_{2}^{\star}$ and ${y}_{1}^{\star}\in A \backslash\left\{y_{1}\right\}$.
On the one hand, by Proposition \ref{single-mc}, we have
\begin{equation}\label{I2-mc}
mc\left(y_{2}, I_{2}\right)\le 3 mc\left(y_{2}^{opt}, I_{2}\right)\le 3 mc\left({y}_{1}^{\star}, I_{2}\right).
\end{equation}
On the other hand,
\begin{eqnarray}
mc\left(y_{1}^{\star}, I_{2}\right) &=&\max _{i \in N_{2}}\left|y_{1}^{\star}-x_{i}\right| \leq \max _{i \in N_{2}}\left|y_{2}^{\star}-x_{i}\right|+\left|y_{1}^{\star}-y_{1}\right| \\
& \leq& \max _{i \in N_{2}}\left|y_{2}^{\star}-x_{i}\right|+\left|y_{1}^{\star}-\mathrm{lt}\left(\mathbf{x}_{N_{1}}\right)\right|+\left|\mathrm{lt}\left(\mathbf{x}_{N_{1}}\right)-y_{1}\right|\\
& \leq& m c\left(y_{2}^{\star}, I_{2}\right)+2m c\left(y_{1}^{\star}, I_{1}\right) \label{I12-mc}
\end{eqnarray}

Combining Eq. (\ref{I2-mc}) and Eq. (\ref{I12-mc}), we have
\begin{equation}\label{I221-mc}
mc\left(y_{2}, I_{2}\right) \leq 3 mc\left({y}_{2}^{\star}, I_{2}\right)+6 mc\left({y}_{1}^{\star}, I_{1}\right)
\end{equation}

Thus, by Eq. (\ref{I1-mc}) and Eq. (\ref{I221-mc}), it holds that
\begin{eqnarray}
mc(f,(\mathbf{x}, \mathbf{p}))&=&\max \left\{mc\left(y_{1}, I_{1}\right), m c\left(y_{2}, I_{2}\right)\right\} \\
& \leq& \max \left\{3 m c\left({y}_{1}^{\star}, I_{1}\right), 3 mc\left({y}_{2}^{\star}, I_{2}\right)+6 mc\left({y}_{1}^{\star}, I_{1}\right)\right\} \\
& \leq& 9 m c(O P T,(\mathbf{x}, \mathbf{p}))
\end{eqnarray}

Above all, \textit{Mechanism 4} has an approximation ratio of at most $9$.
\end{proof}

\end{document}